# Economic Policy Uncertainty: A Review on Applications and Measurement Methods with Focus on Text Mining Methods

Fatemeh Kaveh-Yazdy [1]

Department of Computer Engineering,
Yazd University, Yazd, Iran
fkavehy@stu.yazd.ac.ir
(ORCID ID: 0000-0003-3559-5510)

Sajjad Zarifzadeh

Department of Computer Engineering,
Yazd University, Yazd, Iran
szarifzadeh@yazd.ac.ir

## Abstract

Economic Policy Uncertainty (EPU) represents the uncertainty realized by the investors during economic policy alterations. EPU is a critical indicator in economic studies to predict future investments, the unemployment rate, and recessions. EPU values can be estimated based on financial parameters directly or implied uncertainty indirectly using the text mining methods. Although EPU is a well-studied topic within the economy, the methods utilized to measure it are understudied. In this article, we define the EPU briefly and review the methods used to measure the EPU, and survey the areas influenced by the changes in EPU level. We divide the EPU measurement methods into three major groups with respect to their input data. Examples of each group of methods are enlisted, and the pros and cons of the groups are discussed. Among the EPU measures, text mining-based ones are dominantly studied. These methods measure the realized uncertainty by taking into account the uncertainty represented in the news and publicly available sources of financial information. Finally, we survey the research areas that rely on measuring the EPU index with the hope that studying the impacts of uncertainty would attract further attention of researchers from various research fields. In addition, we propose a list of future research approaches focusing on measuring EPU using textual material.

## Keywords

Monetary Policy, Economic Policy Uncertainty (EPU), Text Mining, Social Media, News Mining.

## JEL Classification

C53, C38, A13, O38, H50.

---

[1] Corresponding author at: Department of Computer Engineering, Yazd University, Yazd, Iran. PO Box: 98195 – 741.
E-mail addresses: fkavehy@stu.yazd.ac.ir (*F. Kaveh-Yazdy*), szarifzadeh@yazd.ac.ir (*S. Zarifzadeh*).

# 1 Introduction

One of the requirements of the economic growth is investment. Investors make their decisions based on the information collected from different sources, such as news wires, social media, analysts' forecasts and managers' disclosed information. In fact, investors receive their required information from a chain of sources which are mixed with noise and uncertainty.

Knight (1921) coined the term "economic uncertainty" (EU) to define the situation in which decision-makers are unable to predict the future outcomes of their decisions (Castelnuovo et al., 2017). He divided the future events based on their probability distribution. Events with a known probability distribution are called risky events, and that of with unknown distribution are uncertain events. Uncertainty can affect economy via two channels: (1) investment, (2) consumer demand. Bloom (2009) indicated that higher uncertainty urges the investors and firm owners to put a delay on their investment plans to decrease irreversible costs, a.k.a. wait-and-see policy. While uncertainty hinders the investment, Caldara et al. (2016) identified uncertainty as well as financial shocks as relevant drivers of the US business cycles. Under higher uncertainty, individuals and firms hold cash and take precautionary actions (Caballero, 1990; Demir & Ersan, 2017; Feng et al., 2019; Im et al., 2017; Phan et al., 2019). Saving money decreases costs and thereby consumption expenditures. Cutting the investment and consumption indirectly exerts the unemployment rate, leading to the post-recession economic recovery delay (Perić & Sorić, 2018). Following the 2008 recession, economic uncertainty attracts a great deal of attention in economy and finance communities due to its ability to predict future recessions, e.g., Ercolani and Natoli (2020) showed that macroeconomic indicators, financial uncertainty and yield curve slope can be used to predict the US recessions. With respect to the context of different economic problems, various instances of uncertainty measures, such as economic policy uncertainty (S. R. Baker et al., 2016), health care policy uncertainty (Cheng & Witvorapong, 2019), and government uncertainty (Saltzman & Yung, 2018) are defined.

Among uncertainty measures, the economic policy uncertainty (EPU) index is widely studied. Governments control tax, income, investment, as well as health care and environmental issues through policy-making. Thus, uncertainty regarding the financial outcomes of policies yields economic uncertainty, e.g., changes in the US government's policies play a more important role in economic uncertainty's alteration than the negligible economic activities (Saltzman & Yung, 2018). Furthermore, increasing the uncertainty in big economies affect not only themselves but also influences economies that are linked to them; for example, changes in the US EPU adversely affect the economic growth of Mexico (Alam & Istiak, 2020) and China (Huang et al., 2020; Li et al., 2020; Li & Peng, 2017).

There are three groups of methods adopted to measure economic uncertainty. The first group uses economic/financial parameters, e.g., CBOE[2] volatility index (VIX)[3], to estimate the economic uncertainty. The second group uses text mining methods to extract the information from various textual resources such as news media (Baker et al., 2016), monetary policy documents (Lee et al., 2019), and tweets (Altig et al., 2020). Text mining methods use publicly available textual resources and are benefitted from recent advances in big data processing frameworks. The third group of methods uses miscellaneous sources of data to estimate the uncertainty. Using these methods shed light on studying the economies with little transparency by taking into account a massive amount of public information. We should note that investigating economic systems using various uncertainty measures reveals different aspects of economic uncertainty (Altig et al., 2020). In this article, we aim to address the following topics,

- Importance of measuring economic uncertainty

---

[2] Chicago Board Options Exchange (CBOE)
[3] https://www.cboe.com/tradable_products/vix/

We review a concise list of domains that are influenced by the economic uncertainty or interact with it. These domains are widely scattered and range from socio-economic (de Bruin et al., 2020; Vandoros et al., 2018, 2019) to climate change (Golub, 2020; Guo et al., 2019; Lecuyer & Quirion, 2019) and tourism (Demir & Gozgor, 2018; Gozgor & Demir, 2018; Işık et al., 2020). To the best of our knowledge, there is no study that has reviewed the influential domains of the economic uncertainty generally or economic policy uncertainty specifically. In this context, we focus on various domains to introduce the new horizons for interdisciplinary researches linking text-based EPU indices to text-based measures that are able to indicate the impacts of economic policy uncertainties.

- Major text mining methods adopted for measuring economic policy uncertainty

  While market volatility indices measure the uncertainty level directly, there are text-based indices that measure the uncertainty level perceived by the news audiences. Although Baker et al. (Baker et al., 2016) introduced a popular EPU measure for the first time and formularized it, but diverse EPU measures have been innovated thus far. These measure are varied in the method and raw data they use which give them the ability to discover domain and diversity of the influences of economic policies. To answer this question, we discuss major EPU measures and weight their pros and cons.

The rest of the paper is organized as follows. Economic policy uncertainty measures are presented and compared in Section 2. Section 3 focuses on major groups of the text mining-based EPU measures. Section 4 briefly reviews applications of the EPU measurement, and finally, we conclude our article and suggest a concise list of future works in Section 5.

## 2    Economic Policy Uncertainty Measures

Economic uncertainty measures can be divided into three categories: (1) financial measures, (2) textual measures, and (3) miscellaneous methods. The first class uses financial parameters and forecasts to measure the consequences of uncertainty in financial systems directly. The second class measures it by taking into account the level of uncertainty that can be perceived by the investors and decision makers. This class of methods measures economic policy uncertainty inferred from the output of the monetary policies reflected in news wires and social media. We should note that there exists a small number of methods that are occasionally used for measuring uncertainty. Figure 1 shows the main categories of EPU measures grouped based on type of their input data. In the following subsections, we discuss each group of methods.

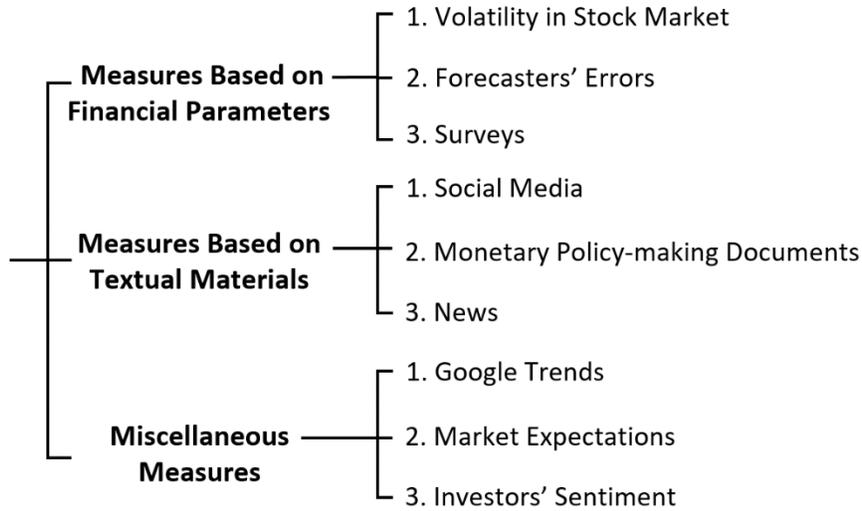

*Figure 1: Categories of economic uncertainty and economic policy uncertainty measures grouped based on type of their input data.*

## 2.1 Uncertainty Measures based on Financial Parameters

### 2.1.1 Volatility-based Measures

Pastor and Veronesi (2012) defined two types of uncertainty, that are (1) political uncertainty, (2) impact uncertainty. The former addresses the public uncertainty about the decisions that government makes and the latter describes the uncertainty about after-effects of a political change. They showed that under policy uncertainty, the stock market returns are negative and the magnitude of the negative returns relies on the level of uncertainty. These findings was consistent with Bloom's (2009) results that suggested uncertainty and real activity have a counter-cyclical relation that can be proxied by stock market. Uncertainty measures that follow volatility in stock market and GDP have been devised based on these findings. These measures follow the realized or implied fluctuations in the stock market and demonstrate the uncertainty. CBOE volatility index (VIX for short) is the most famous stock market volatility index computed and published by the Chicago Board Options Exchange and is also referred as fear index. VIX is computed as follows,

$$VIX = \sqrt{\frac{2e^{r\tau}}{\tau}\left(\int_0^F \frac{P(K)}{K^2}dK + \int_F^\infty \frac{C(K)}{K^2}dK\right)} \tag{1}$$

where $\tau$ is the number of average days in a month ($\tau = 30$), $r \geq 0$ is the risk-free rate, $F$ is the 30-day forward price in the S&P 500. $P(K)$ and $C(K)$ are prices for puts and calls with strike and 30 days to maturity. In fact, VIX is a forward predictor of volatility of S&P 500 stock returns in the next 30 days ( Baker et al., 2016). VIX is widely used in economic studies to investigate the impact of uncertainty (Al-Yahyaee et al., 2019; Chung & Chuwonganant, 2014; Völkert, 2015). Another instance group of volatility-based measures includes dispersion measure. These methods are designed based on the fact that the higher the uncertainty, the more disperse the parameter values are, e.g., dispersion in firm-level as well as industrial-level earnings and total factor productivity can be used as uncertainty measures (Bloom et al., 2018).

### 2.1.2   Forecasters' Errors Measures

The idea behind forecasters' error measures is that, if the forecasters are uncertain about the economic situation, their forecasts are not accurate and their errors rise. Three groups of uncertainty measures are computed using forecasts' errors. The first group uses the disagreement from point forecasts as an uncertainty measure. For instance, Lahiri and Sheng (Lahiri & Sheng, 2010) decomposed the disagreement values into common and idiosyncratic shocks and showed that the disagreement in forecasts can be reliably used as an uncertainty proxy. The second group defines an uncertainty measure that is the equal-weight mean of the individual perception of uncertainties. Jurado et al. (Jurado et al., 2015) developed an uncertainty estimator computed by a weighted average of individual uncertainties of forecasters. They compared their proposed weighting schema with a base-case equally-weighted measure and showed that their predicted uncertainty shocks are coincides with recessions, wars and financial downturns. The third group uses the distribution of the errors to understand the dispersion of the forecasts. Subsequently, the more the dispersion, the higher the certainty. Rossi and Sekhposyan (2015) constructed a set of indices to measure uncertainty with respect to the distribution of forecasters' errors. Indeed, they tried to find that to which extend the realization of a macroeconomic variable is unexpected for the forecasters. Gao et al. (Gao et al., 2019) targeted economic activity uncertainty (EAU) and inflation uncertainty (IU) by defining an uncertainty measure that estimates the unforecastable component of output and inflation. Their results showed that their EAU and UI indices can explain the UK's cross-section of stock market returns.

### 2.1.3   Survey-based Measures

Survey-based methods are interconnected with measures defined based on forecasters' errors. This is because surveys collect analysts' points and forecasts on financial parameters such as GDP, stock prices, and inflation to compare predicted values with their realized values to estimate the uncertainty. Poncela and Senra (2017) developed an uncertainty measure by adopting the European Central Bank Survey of Professional Forecasters (ECB-SPF). The ECB-SPF has a panel of one hundred professional forecasters providing long-term predictions for GDP, inflation, and the unemployment rate in Europe. They suggested a mixed weighting system which uses a principal component analysis to select $r$ principal components ($r < n$, where $n$ is the total number of uncertainty parameters) and then forecasts the values using a panel regression. They evaluated the accuracy of the predicted GDP and inflation under selected uncertainty components and found that the forecasting errors entered the model as surprise parameters substantially increase the prediction performance. Although survey-based measures are frequently used in research works (e.g., (Bachmann et al., 2013; Claveria, 2020; Gao et al., 2019; Glas & Hartmann, 2016; Leduc & Liu, 2016)), Claveria (2020) showed that the inflation uncertainty estimated by the forecasters is different. According to the results, disagreements in expansionary are lower than that of in contractionary periods, which means forecasters' estimations can be varied depending on their overall feelings.

## 2.2   Uncertainty Measures based on Textual Materials

### 2.2.1   Social Media-based Measures

Baker et al. (2020) introduced a novel Twitter-based Economic Uncertainty (TEU). Then, they adopted TEU as well as news-based and financial uncertainty measures to study the impacts of COVID-19 pandemic on the US and UK's economy (Altig et al., 2020). Results showed that all of uncertainty measures jump up during the pandemic, but measures behave differently in their rising window. These differences include variations in the number of rising folds, e.g., uncertainty of the UK's sales growth rises 100% while the forecasters' disagreement about the UK's GDP growth jumps 20-fold. Moreover, the rising-recovery paths are also different, e.g., stock market volatility rises in later

February and the increases recedes in late March. On the other hand, measures that are affected by the unemployment surges peaked later. They indicated that although all uncertainty measures are useful and are able to demonstrate aspects of uncertainty, the Twitter-based measure is more credible to show households' perception of uncertainty. Barrero and Bloom (2020) used news-based and Twitter-based uncertainty measures to investigate the recovery after COVID-19 pandemic and future of remote working. They inferred that the reallocation is required as COVID-19 enforced the cross-firms and industries to move the capital and labor which is resulted in higher uncertainty. Furthermore, remote working regulatory retards hiring for firms concerning about the onboard personal training barriers. In the same manner, Becerra and Sagner (2020) investigated the impacts of COVID-19 pandemic on the exchange rate in Chile using two Twitter-based uncertainty measures and found that the peaks of these measures are rising a week after the COVID-19 outbreak in Chile which is consistent with Baker's et al. (Baker,et al., 2020) findings.

### 2.2.2    Measures based on Monetary Policy-making Documents

Monetary policy-making documents present the economic policies directly. Therefore, tracking the changes in monetary policies reveals the uncertainty hidden behind the decisions. In this way, Saltzman and Yung (2018) focused on the polarity of sentences addressing the economic policies in the Federal Reserve Beige Books. They computed the overall polarity of every Beige Book by differencing the total number of positive sentences mentioned in an economic policy from negative ones. The total polarity of a book is normalized by its total number of words. Generated time series are scaled by 10,000 and aggregated on a quarterly basis. Although their proposed method is reasonably accurate and can deal with a smaller set of textual data, it is not applicable to economies that do not have Beige Books-like commentaries. Lee et al. (2019) designed a pipeline which receives the minutes of the Bank of Korea (BoK) monetary policy board and extract two uncertainty measures. The first method uses an economic field-specific dictionary to extract financially credible sentences and represents them with ngram2vec. The second one constructs a sentiment index based on the market analysis approaches. The total number of the positive-toned sentences is differenced from that of the negative-tone sentences. Then, they showed that their measure outperforms the news-based and volatility-based measures.

Hansen et al. (2017) analyzed the minutes of the three central banks, that are the Federal Reserve, the Bank of England and the European Central Bank, using topic modeling methods to investigate the relation between transparency in central banks' communication and deliberation level of monetary policy-makers. Their findings suggested that the economic uncertainty is associated with the fiscal and risk topics in minutes. Binette and Tchebotarev (2019) tried to detect the changes in economic conditions and its direction using the Bank of Canada's Monetary Policy Reports (MPR). They hypothesized that the lexical innovation in MPRs reveals the uncertainty hidden behind the monetary decisions. Based on this hypothesis, they utilized the Word Mover Distance (WMD) (Kusner et al., 2015) to track changes in the MPRs. Then, they proposed Universal Language Model Fine-tuning (ULM-FiT) to determine the tone (sentiment) of MPRs' sentences. The tone scores are aggregated by differencing the number of positive sentences from negative ones in each MPR. Finally, time series of uncertainty are constructed by summing up the WMD scores and tones of MPRs. Results of investigations showed that peaks of the series coincide with the events associated with high uncertainty, such as 9/11 and 2008 global financial crisis.

### 2.2.3    News-based Measures

News-based EPU measures are designed to measure the uncertainty that might be perceived by society via information provided by news media. Baker et al. (2016) introduced a news-based EPU measure for the first time. They searched the archive of ten US newspapers and selected the articles that included terms "uncertainty" or "uncertain"; "Economy" or "Economic", and a set of terms conveying the policy-making meaning, e.g., "legislation", "Congress", and "regulation". The raw number of articles that pass the conditions in each newspaper over each month is normalized by the total number of articles in the corresponding newspaper and month. The normalized monthly series are standardized to unit deviation and then divided by the monthly averages across newspapers. The normalized series are scaled to a mean of 100. Baker et al. (2016) computed the economic policy uncertainty index (henceforth, BBD[4] index) for twelve major economies, including the United States, China, Russia, Germany, and Japan[5]. Moreover, they enlisted eight subtopics in articles demonstrated the sectors of uncertainty in economic policy, as (1) fiscal policy, (2) monetary policy, (3) health-care, (4) national security, (5) regulation, (6) sovereign debt and currency crisis, (7) entitlement programs, (8) trade policy. Baker et al. (2016) in collaboration with different research teams have shared the EPUs of 26 countries since 2012 in a public website under the *www.policyuncertainty.com* domain.

Azqueta-Gavaldón et al. (2020) used news collected from 12 newspapers in Germany, France, Italy and Spain. They pre-processed the articles and applied a topic modeling method. Identified topics are grouped under eight sectors of uncertainty with respect to their top keywords. Decomposing the uncertainty to its sectors let them assess the impact of each sector in countries separately. Manela and Moreira (2017) constructed a news-based uncertainty proxy called NVIX using front-page articles of the Wall Street Journal (WSJ). They extracted the likelihood of the monograms and bigrams frequently appeared in the body of articles and adopted a linear regression which takes likelihoods as features to predict the VIX index. Indeed, the NVIX measure is a news-based predictor of the VIX uncertainty. Furthermore, the weights of the features in the regression model uncover the n-grams that are associated with uncertainty sources, such as wars and rare disasters (Manela & Moreira, 2017).

Xie (2020) developed an EPU measuring model named Wasserstein Index Generation model (WIG) which feed by the word-embedding vectors of the New York Time's articles. This method tries to label documents with topics that are extracted based on the distribution of the words. At the end, they applied a Singular Value Decomposition (SVD) to the matrix of words-topics and estimate a single document-wise score that yields the corresponding uncertainty level. The series of normalized monthly-aggregated scores are used as uncertainty index series. The reported results showed that this uncertainty proxy is highly correlated with the EPU measure proposed by Baker's et al. (2016) and the Azqueta-Gavaldón's measure (Azqueta-Gavaldón, 2017).

The real option theory (Bernanke, 1983) indicated that the delay policy could be considered as an option for investors in situations with higher levels of uncertainty and negative sentiments. With respect to this theory, researchers tried to design a sentiment proxy to measure uncertainty indirectly. In this way, Tetlock (2007) hypothesized that the pessimism in media can be used as a measure of downward marker for market prices. He counted the negative terms in one of the most popular columns of the Wall Street Journal. Results showed that unusual high or low values of his proposed measure coincide with high market trading volume. Among the different EPU measures, news-based measures are mostly used in economic studies to investigate the impacts of uncertainty on firms and households decision-making.

## 2.3   Miscellaneous Measures

Although major methods can be categorized into famous measures using financial parameters and indices developed by adopting textual materials but there is an extra class of methods that are rarely defined and used. Jurado et al. (2015) constructed a model-based EPU measure which combines more than one hundred micro-economic variables and financial indicators to benefit from the advantages of measuring uncertainty in various sectors. Binge and Boshoff (2020) combined a business tendency survey-based uncertainty measure with four other uncertainty measures (selected from financial-based and text-based groups) using principal component analysis (PCA) model. In the similar manner, Dai et al. (2020) consolidated a global economic policy uncertainty index by combining the national EPUs of twenty primary economies. They generated a panel data of the EPU indices and adopted the PCA to reduce the dimension of the data to one column that is the global EPU (GEPU). They compared their GEPU with the measure proposed by Davis (2016) which was a GDP-weighted average of national EPU indices for twenty economies (named GDP-GEPU)[6]. Dai's et al. (2020) GEPU measure correlates with the GDP-EPU and follows the stock market volatility. As a step in developing a world-scale EPU measure, Ahir et al. (2019) defined the World Uncertainty Index (WUI) concept and computed this index for 143 countries with a population of greater than two millions. World Uncertainty Index is a series of the frequency of the word "uncertainty" repeated in the quarterly country report of the Economist Intelligence Unit (EIU). The series are normalized by the total number of words in reports.

---

[4] BBD comes from initials of Baker, Bloom, and Davis (S. R. Baker et al., 2016)
[5] Available at www.policyuncertainty.com

Hamid and Heiden (2015) extracted Google Trends (GT)[7] daily time series of the term "Dow" which is correlated with the "Dow Jones" searches and found that the GT time series are correlated with the stock volatility uncertainty but lagged two days behind the volatility uncertainty. Castelnuovo and Tran (2017) constructed two uncertainty measures for the United States and Australia by leveraging the uncertainty-related keywords appeared in the monetary policy-making documents such as Federal Reserve's Beige Book and the Reserve Bank's Monetary Policy Statement of Australia as search terms. The aggregated GT series of these terms are used in developing a Google Trends-based Uncertainty (GTU) measure which is positively correlated with financial and text-based uncertainty measures. Donadelli and Gerotto (2019) built four uncertainty indicators using GT series to cover the health-, environmental-, security-, and political-related uncertainties. Scotti (2016) defined a surprise index based on the market participants' expectations about the economy with respect to the Bloomberg forecasts. Da et al. (2015) aggregated searches related to households, such as recession, global crisis, bankruptcy and unemployment to construct a measure that shows the investors' sentiment. This measure, named FEARS, can be used to predict short-term return reversals and temporary increases in volatility. Their results follow the hypothesis that the stock return is predictable by taking into account the investors' sentiments. Guo et al. (2021) and Zhang (2019) reported negative correlation with investor sentiments and EPU.

## 2.4    Comparison of EPU Measures

In this section, we briefly compare major EPU measures and review their advantages as well as disadvantages. For the sake of space, we summarized the characteristics of the measures in Table 1.

*Table 1: Comparison between groups of economic policy uncertainty measures with respect to the input data type.*

| Data Type | Idea | Characteristics | Advantage(s) | Disadvantage(s) |
|---|---|---|---|---|
| Stock Volatility | The more volatile the stock market, the higher uncertainty is realized. | • Limited horizon (e.g., 30-day)<br>• Shows economic uncertainty in short-run (e.g., 30-day) (S. R. Baker et al., 2016)<br>• Defined based on S&P 500 index | • As a fear index demonstrates social economic concerns[8] and sentiments<br>• Predicts the future uncertainty (moves forwards) | • Stock market volatility changes over time even if there is no change in economic uncertainty (Jurado et al., 2015)<br>• Implied uncertainty partly demonstrates the economic situation (Girardi & Reuter, 2017) |
| Forecasters' Errors | The more the forecasters' error, the more volatile the stock market is and the higher the uncertainty is realized. | • Based on forecasts of a few number of forecasters of stock market returns.<br>• Uses different aggregation policy, such as weighted mean | • Demonstrates the uncertainty realized by the professional analysts<br>• Predicts the future uncertainty (moves forwards) | • Limited number of board members<br>• Analysts' realized uncertainty might not be the same as the uncertainty realized by the non-professional investors (e.g., households)<br>• Expensive and time consuming |
| Survey-based | The more diverse the forecasts, the higher the implied uncertainty. | • Based on forecasts of boards of financial experts forecasting financial parameters, such as GDP and inflation<br>• Short to middle size window size<br>• Collected by central banks | • Demonstrates the uncertainty realized by the professional analysts<br>• Predicts the future uncertainty (moves forwards)<br>• Due to formality of providers, it can be used as a reliable resource for policy-makers<br>• More accurate than forecasters' error –based measures | • Limited number of board members<br>• Analysts' realized uncertainty might not be the same as the uncertainty realized by the non-professional investors (e.g., households)<br>• Expensive and time consuming |

---

[8] https://www.cboe.com/tradable_products/vix/

| | | | | |
|---|---|---|---|---|
| Social Media | The more negative and uncertain the comments are; the more economic uncertainty is realized. | • The data is provided by collecting Tweets, Facebook posts and other social media posts<br>• Requires uncertainty implication keywords<br>• Requires sentiment (tone) detection tools<br>• Unlimited time horizon (from past to present) | • Demonstrates the EPU realized by the public community and households<br>• Publicly-available data<br>• Can be used to study economies with little to no transparency | • Diverse<br>• Might show inconsistent behavior in comparison with the uncertainty measures computed based on financial parameters<br>• Require mining massive amount of data<br>• Sensitive to the accuracy of sentiment analysis methods used to annotate the data |
| Monetary Policy-making Documents | Changes in tones and topics in monetary policy-making documents demonstrates the policy-makers' uncertainty. | • Uses publicly-available formal economic policy-making documents<br>• Requires sentiment (tone) detection tools<br>• Track changes in tone<br>• Unlimited time horizon (from past to present) | • Measure the economic policy uncertainty directly<br>• The uncertainty that realized by the analysts, news reporters and authors does not intervene the implied uncertainty | • Applicable when the policy-making documents and minutes are published publicly.<br>• Require mining massive amount of data<br>• Sensitive to the accuracy of sentiment analysis methods used to annotate the data |
| News | The proportion of the news economic article includes policy-making keywords and implied uncertainty shows the magnitude of the EPU realized by the society. | • Uses news articles<br>• Requires standard list of terms related to policy-making, uncertainty and financial concepts | • Publicly available data resources<br>• Track the uncertainty in the source that is used by small-/large-scale investors | • Sensitive to the keyword list<br>• Hard to generalize for languages other than English<br>• Require mining massive amount of data |

# 3 Text Mining Methods for EPU Measurement

## 3.1 Keyword-based Methods

Although the BBD computation is introduced in section 2.2.3, the keyword selection procedure has not been discussed. Baker et al. (2016) sampled a set containing 12,009 articles from 1900 to 2012 and hired research assistant to label them. In the next step, terms that are frequently occurred in the positively-labeled documents and associated with the policy-making are extracted by auditors. Then, a set of 32000 terms that are co-occurred with the extracted terms in at least four documents are selected and their distribution in positively- and negatively-labeled documents are generated. Finally, $P$ terms that minimize the gross error rate[9] are chosen to be used as policy-making indicators. The uncertainty measure is calculated by counting the number of articles containing "uncertainty", "Econom\*" and policy-making terms in each newspaper. The number of articles are normalized by the total number of articles in each month, then standardized to unit deviation and finally aggregated by the monthly averages of the values for other newspapers.

In the same line, Movsisyan (2018) estimated the BBD index for Armenia and showed that this index is consistent with several macroeconomic indicators such as country risk premium, exchange rate, and bank lending tightening. Luk (2018) focused on Macao's game and gambling economy and adopted a repository of local newspaper articles to compile the Macao's BBD index. He showed that the index is correlated with the gross gaming revenue of Macao and has a negative impact on unemployment growth. Luk et al. (2020) constructed BBD index for Hong Kong for the period of 1998 to 2016 and showed that rises in EPU results in unemployment increases and lower investment.

Baker et al. (2016) estimated China's uncertainty index using one English newspaper published in Hong-Kong to liberate their analysis from China's censorship. Applying this restriction on news resources raises several issues. First, any change in the viewpoint of the editorial team of the newspaper leads to significant changes in the index. Second, Hong-Kong's economic news articles have a higher priority for the newspaper editorial than China's economic policy; therefore, China's economic policy might not be covered by this newspaper in detail. Third, the changes in the index do not follow some of the Chinese macroeconomic variables. Forth, their news database is sparse and does not contain enough articles to follow daily economic changes or drill-down on trends to find the reason behind spikes. Huang and Luk (2020) computed the BBD index using ten major Chinese newspapers published in mainland China. They showed that their index follows the findings upon the relation between EPU and other economic variables in the US economy. Several research groups adopted Baker's et al. (2016) method to compute uncertainty measures for their preferred countries. In this way, Hołda (2019) estimated EPU of Poland, Ghirelli et al. (2019) compiled Spanish news articles to develop EPU for Spain, and EPU of Japan and Brazil are computed by Arbatli et al. (2017) and Ferreira et la. (2019), respectively.

Above mentioned studies computed the BBD index using search terms in their target languages. On the other hand, Braun (2020) adopted the policy-making and uncertainty keywords to compute his uncertainty measure in an innovative way. He indicated that firms' decisions and their outcomes are highly dependent. He deduced that the impacts of uncertain policies on different firms could not be aggregated while different sectors of the economy have their own dynamics. Braun (2020) defined his firm-level EPU by splitting the share of sentences of the 20F reports addressing the economic policy-making from the share stated uncertainty. Finally, the following EPU index is calculated by multiplying the number of sentences of each share.

---

[9] Gross Error Rate = No. of False Positive Cases + No. of False Negative Cases

## 3.2    Topic Modeling

There exist different text mining methods that are alternatively used for EPU estimation. Azqueta-Gavaldón (2017) exploited Latent Dirichlet Analysis (LDA) to estimate topic-specific EPU of the US. LDA model eliminates the requirement for labeled data. Yono et al. (2020) adopted supervised LDA (sLDA) to compute the EPU index for Japan. Their model uses VIX index and numerical signal to supervise the LDA model which is trained without labeled data. Rauh (2019) used LDA topic modeling to compute Canada's EPU indices at the regional level. He grouped news with respect to the region they addressed and applied a topic modeling approach to cover the diversity of policies made in different areas of Canada.

## 3.3    Methods using Word Embedding

Baker et al. (2016) selected policy-related terms manually. Braun (2020), Huang and Luk (2020), Hołda (2019) as well as many other researchers have used keyword set to expand the three major concepts, that are uncertainty, economic, and policy. keyword-based methods are built upon experts' knowledge and can be biased by their interpretation. Moreover, the synonymy relation between terms is context-independent. Azqueta-Gavaldón (2017) and Yono et al. (2020) adopted the LDA topic modeling, which does not require an initial keyword set; however, an expert must inspect topics and assign them to their corresponding uncertainty group. In recent studies, word embedding representations are widely used in semantic analysis, while they are able to discover synonymies approximately. Tsai and Wang (2014) employed the dictionary developed by Loughran and Mcdonald (2011) and annotated the dictionary's words by their Part-of-Speech (POS) tags. Tagged words are expanded by their top-20 most similar terms using the Continuous Bag-of-Words (CBOW) model. Rekabsaz et al. (2017) suggested term filtering based on a pre-defined threshold applied to the cosine similarity of the words relevant to uncertainty concepts. Theil et al. (2020) trained two-word embedding based models called industry-specific and industry-agnostic word2vec on 10-k corpus. They expanded Loughran and Mcdonald (2011) words by the top-20 nearest neighbors using both word embedding models and then retrieved the economic-related articles. Keith et al. (2020) expand the Baker et al. (2016) keywords with their five nearest neighbors which are collected via applying cosine distance to GLOVE word embedding vectors (Pennington et al., 2014). Their results showed that the BBD computed using expanded keyword set have higher correlation with VIX measure. Kaveh-Yazdy and Zarifzadeh (2021) proposed an unsupervised word embedding-based EPU measure that expands the keywords using cosine distance between vectors of the terms and represents the documents in a plain tri-axial system. This method utilizes does not determine semantic similarity thresholds strictly. Instead, it balances the impacts of the documents by adopting a weighting schema based on the triangle's surface representing the document. In fact, this method summed up the impacts of the documents included the EPU-related terms. Furthermore, this Omni-language method is not sensitive to the keyword richness language.

## 3.4    Document Classification Methods

Tobbak et al. (2018) annotated a set of labeled documents as seed and trained a binary Support Vector Machine (SVM) model to label EPU/Not-EPU-relevant documents. This model is applied to the remained part of their dataset. Then, the documents labeled with a higher level of uncertainty are passed through a pool-based active learning algorithm with uncertainty sampling to optimize the classifier. Keith et la. (2020) sampled 2531 labeled documents from Baker's et al. (2016) corpus. This set was divided into1844 documented from 1985-2007 as training set and 687 documents from 2007-2012 as test set. They cleaned and tokenized the documents to generate a dictionary of terms. Then, a logistic regression classifier which uses the bag-of-word representation is trained and used to label test document. They denoted that the precision of their classifier does not beat the word embedding-based labeler which expands the Baker's et al. (2016) keywords with their five nearest neighbors (mentioned in 3.3).

# 4    Applications of Economic Policy Uncertainty Measurement

Studies around the economic uncertainty generally and economic policy uncertainty specifically can be grouped into two major groups. The first group includes the EPU estimation methods. The second group of EPU-related studies covers the impacts of the EPU on different domains such as economy, behaviors of investors, climate change, and decision making. Moreover, EPU has interaction with parameters such as geopolitical risk which means it can be influenced by risky events such as terror attacks and wars. Although there exist publicly-available sources of textual information, the interconnections between the economic uncertainty and socio-economical parameters have been understudied. In the following subsection, we briefly discuss the primary areas that are affected by the uncertainty with the hope that studying the impacts of uncertainty would attract further attention of the researchers from various research fields.

## 4.1    Health

Economic growth affects health-care in two channels (Cheng & Witvorapong, 2019), (1) rising health service demands, and (2) increasing investment in health-care. Increases in wages raise the demand for extra health-care services and grows up the health expenditures (Cheng & Witvorapong, 2019). On the other side, governments and health-care management initiatives invest on health-care services directly or indirectly through providing incentives for firms (Bloom et al., 2019). Consequently, governments through the policy-making shape the health-care service delivery and health insurance markets (de la Maisonneuve et al., 2017). Then, when firm owners are concerned about the investment returns due to uncertainty around health-care regulations, they cut the investment (Kang et al., 2014). Subsequently, the health-care policy uncertainty would be the source of changes in health-care expenditures and life quality (Baker et al., 2016; de la Maisonneuve et al., 2017; Jakovljevic et al., 2017). Cheng and Witvorapong (2019) used the US's Health-Care Policy Uncertainty (HCPU) measure[10] to investigate the economic consequences of HCPU shocks. They showed that under a higher health-care uncertainty, households and firms cut their expenses and stop investing, respectively which is consistent with the "wait-and-see" policy. Furthermore, uncertainty shocks imposed by new policies decrease the health-care and general inflation.

We should remind that the relation between health and economic uncertainty is bi-directional. On March 1, 2020 the World Health Organization (WHO) announced COVID-19 outbreak as a pandemic. Since then the impacts of the uncertainty caused by the COVID-19 pandemic have been widely studied. In contrast with the cases that study the impacts of health-care policy uncertainty on financial parameters, this group of studies aimed at targeting the uncertainty caused by COVID-19 health crisis. Sharif et al. (2020) investigated the simultaneous impacts of COVID-19 pandemic and oil price volatility on the US economic uncertainty and geopolitical risk. It reveals that the COVID-19 is a greater long-term thread for the economic growth in comparison with the oil price volatility. Moreover, it seems that the COVID-19 affects the US geopolitical risk more than economic uncertainty.

Choi (2020) explored the interdependence and causality between EPU and volatility of eleven S&P 500 index sectors[11] in the US during the COVID-19 pandemic. He discovered that the interdependence degree between the EPU and the sector volatility during the pandemic is higher than that in the global financial crisis (GFC) of 2008. In this way, Altig et al. (2020) compared the EPU of the UK and US before and during the pandemic and concluded that the COVID-19 outbreak leads to a surprising shock in the US stock market which was stronger than the shockes occurred during SARS, Swine Flu and Ebola pandemics. Wang et al. (2020) examined the ability of VIX and EPU measures in forecasting volatility for nineteen equity indices during the pandemic and found that the VIX measure which is also called "panic index" outperforms the EPU index. While the EPU index chnages after publishing and spreading news articles, the VIX is reacted earlier and tends to represent premature reactions.

---

[10] Available at https://www.policyuncertainty.com/categorical_epu.html under the "US categorial EPU Data" section.
[11] Volatility sectors: Communication Services, Consumer Discretionary, Consumer Staples, Energy, Financials, Health, Industrials, Information Technology, Materials, Real Estate, Utilities

## 4.2   Personal Issues and Social Behaviors

Volatility and uncertainty dampen the economic growth and increase the unemployment (Caggiano et al., 2017) which lead to increases in the sense of hopelessness and pessimism about future (de Bruin et al., 2020). A wide range of researches showed that suicide and self-harm extensively increase during recessions and periods with economic hardship and high uncertainty (de Bruin et al., 2020; de la Maisonneuve et al., 2017; Demirci et al., 2020; Jofre-Bonet et al., 2018; Vandoros et al., 2019). The number of motor vehicle collisions in the UK is increasing in periods with higher economic uncertainty due to stress, lack of concentration and sleep deprivation (Vandoros et al., 2018). Higher uncertainty followed by employment security devastate the resilience and increase the anxiety levels as attributed in COVID-19 lockdowns (Brenner & Bhugra, 2020; Godinic et al., 2020; Ruffolo et al., 2021; Sher, 2020). In this way, groups of people with high and middle classes of income surge to initiating long-term life-insurances during recessions and high uncertainty periods to decrease their stress level (Grishchenko, 2019; C.-C. Lee et al., 2021). However, Balcilar's et al. (2020) findings assert that non-life insurance are more sensitive to uncertainty and volatility.

According to the Kalcheva's et al. (2020) observations, high uncertainty leads to deteriorate impulse control which is associated with poor health behaviors, such as drinking and smoking. Economic uncertainty through its labor market channel affects the fertility rate. Even though, a negative relation is determined between fertility rate and the economic uncertainty (Hofmann & Hohmeyer, 2013; Hondroyiannis, 2010). Novelli et al. (2021) discovered that changes in the short-term fertility intention differentiates between male and female workers. Moreover, higher uncertainty may cause undesirable cultural consequences such as individualism and masculinity (Galariotis & Karagiannis, 2020).

## 4.3   Tourism and Traveling

The negative relation between the economic policy uncertainty and the tourism & traveling industry can be hypothetically explained by the "Wait-and-See" policy. In periods with high uncertainty which raises the financial risks, firm owners cut their investment in the tour and traveling industry due to its irreversibility (Demir et al., 2020). On the micro-level, households postpone their traveling plan to decrease their expenditures (Balli et al., 2018). Findings of (Balli et al., 2018; Demir & Ersan, 2018; Dragouni et al., 2016; Ghosh, 2019; Gozgor & Ongan, 2017; Sharma, 2019) suggested that increasing EPU has negative impact on the tourism growth and demand. The size of the impacts relies on magnitude of the EPU shock (Balli et al., 2018), (Liu et al., 2020). Furthermore, uncertainty shocks influence sub-sectors of the tourism industry, e.g. air travel (Tsui et al., 2018), and hospitality (Akron et al., 2020)-(Ozdemir et al., 2021). Although, this results are theoretically expected, but are rather inconsistent with a group of studies reporting positive relations between the EPU and the tourism demand (Nguyen et al., 2020; Nguyen et al., 2020; Wu & Wu, 2020). In this way, Sharma (2019) investigated the relation between the EPU and the tourism demand with respect to the time horizon (short-term and long-term) and reported that this relation is asymmetric. Aharon et al. (2020) addressed this issue by taking into account the impact of sentiment. They concluded that uncertainty measures such as EPU and VIX are not able to predict fluctuations in the tourism and leisure market accurately alone and contributing the consumer sentiment indices improves the performance of the prediction models. Ballie et al. (2018) and Nguyen et al. (2020) indicated that higher uncertainty have distinct impacts on domestic (inbound) and outbound traveling. It appears that the outbound tourism is more sensitive to the negative impacts of the EPU. In addition, the wealth level of travelers shapes the direction of the relation so that the relation between EPU and domestic as well as outbound is negative in higher-income economies and positive in lower-/middle-income economies. We conclude that the time horizon, wealth-level, and the consumer sentiment indices should be included in inspecting the relation between the EPU and growth of the tourism industry to ensure about the accuracy of results.

## 4.4    Real Estate and Housing Market

Housing and construction markets are primary drivers in developing countries such as China and India (Chow et al., 2018), and the housing market is regulated by the laws enacted by the public sectors. Consequently, they are susceptible to policy changes. The impacts of the economic policy uncertainty on the market growth and pricing in these markets have been investigated in various countries along different time periods (Cakan et al., 2019; Choudhry, 2020; Chow et al., 2018; Huang et al., 2020; Jeon, 2018; Kirikkaleli et al., 2020). Results showed that there exists a negative relation between the EPU and house price indices (Anoruo et al., 2017; Choudhry, 2020; Christou et al., 2017, 2019; El-Montasser et al., 2016; Jeon, 2018; Z. Li & Zhong, 2020; Ongan & Gocer, 2017). This unidirectional significantly negative relation enables the models fed by the EPU values to predict the house price indices and house sales accurately (Anoruo et al., 2017; Çepni et al., 2020; Chow et al., 2018; El-Montasser et al., 2016). Even though increasing the uncertainty level leads to lower house prices and a higher risk of losses (Antonakakis et al., 2015; Aye et al., 2019; W.-L. Huang et al., 2020), it drives the house market because investment in construction and house markets hedges against the losses in companion investments (Aye et al., 2019). For example, Kirikkaleli et al. (Kirikkaleli et al., 2020) identified bubbles around the German housing sector in high EPU periods, and Cakan et al. (2019) reported the herding behavior for investment in the housing market of South Africa under high EPU periods.

## 4.5    Energy Consumption and Climate Change

The largest share of energy in the world comes from fossil fuels resulted into carbon dioxide emission (CO2) (Yu et al., 2021). Carbon dioxide as a greenhouse gas accelerates the global warming and drives climate changes. Governments, non-profit organizations and the United Nation adopted the Paris Agreement[12] to control and limit the CO2 emission. The primary step in this way is enacting laws and regulations that enforce the firms to limit their carbon dioxide. While governments shape the energy consumption and carbon emission regulations, the economic policy uncertainty affects the climate change and the CO2 emission rate (Jiang et al., 2019). EPU intervenes in the climate change issues in three ways as follows, (1) the renewable energy investment policies, (2) prices of the non-renewable energy resources, and (3) decreasing the consumption. Investing in renewable energy firms and green innovation companies have been doubtful and policy uncertainty hinders the advancements in these area (Adams et al., 2020; Contreras & Platania, 2019; Fang et al., 2018; Golub, 2020; Sarkodie et al., 2020).

The conventional energy resources (such as oil, gas and coal) are adversely impacted by the global and national policy uncertainties (Lecuyer & Quirion, 2019; Lin & Bai, 2021; Liu et al., 2020). Results showed that higher EPU level inhibits investment in the oil and gas industries globally. Lin and Bin (2021) showed that the influence of the global economic policy uncertainty can be different between the oil exporters and oil importers. According to their results, higher EPU in oil exporter countries decreases the oil price which benefits the oil importer countries and reduces their national EPUs. Even though, the EPU fluctuating in short-run have negative/positive relation with oil price in oil exporter/importer countries, EPU increment in long-run inhibits the energy investment. It is worthwhile to note that the renewable energy resources are not significantly affected by the EPU (Liu et al., 2020). The third channel for EPU intervention is consumption. Financial growth leads to income increment and demand boost; which headed for more energy consumption and CO2 emission (Adams et al., 2020; Sarkodie et al., 2020). Consequently, higher EPU reduces the consumption which is followed by the production decrement and lower $CO_2$ emission.

---

[12] https://unfccc.int/process-and-meetings/the-paris-agreement/the-paris-agreement

## 4.6    Geopolitical Political Risk and Political Polarization

The sets of economic policy uncertainty causes and geopolitical risk (GPR) factors have common members. Studying their interactions and co-movement during shocks and recessions reveals diverse dimensions of events with respect to their outcomes. Moreover, GPR and EPU together rise adverse impacts on economy; however, the negative impacts of the EPU are ahead of the GPR (Caldara & Iacoviello, 2018). Consequently, investigating the impacts of the GPR and EPU in areas affected by sources of tensions is inevitable (e.g., tourism (Jiang et al., 2020), oil market (Lin & Bai, 2021)). In this way, Tiwari et al. (2019) investigated the impacts of the GPR and EPU on tourism and showed that the negative effects of GPR on tourism run in long-term horizons while that of the EPU influence the tourism industry in short-run windows. Kannadhasan and Das (2020) explored the reaction of Asian markets to the GPR and EPU shocks and suggested that each variable has its own influence. That is, the adverse impact of the EPU is stronger than the GPR and covers all quantiles of the stock market returns, while GPR is negatively correlated with returns in lower quantiles and positively-related with higher quantiles. Thereby, GPR and EPU might be affected by the same event or crisis such as wars (Caldara & Iacoviello, 2018), trade-conflicts (He et al., 2020) and pandemics (Sharif et al., 2020).

Political polarization is another area that is linked with by the economic policy uncertainty. Baker et al. (Baker et al., 2020) addressed the EPU around the political elections in 20 countries including the US presidential elections. They showed the EPU values are higher near the elections and spikes are sharper when the polarization maximizes. In another study, Baker et al. (Baker et al., 2014) explored two explanations for the US rising uncertainty and concluded that the main reason behind the higher uncertainties since 1960s is political divisive and polarized atmosphere which divide the political power required to enact economic policies. Political elections tend to affect the EPU due to their impacts on policy alteration; whereas, referendums cause spikes in the EPU values. For example, the EPU level of the UK increased to its highest level after Brexit referendum (Azqueta-Gavaldon, 2017; Belke et al., 2018; Nilavongse et al., 2020) resulted in massive deprecation in British pound exchange rate. Unexpected acceptance of a referendum in February 2014 in Switzerland increased the country's EPU and resulted in investment reversal.

## 4.7    Micro-/Macro-economic

The most important research areas that are tightly linked with economic policy uncertainty measures are micro-/macro-economic decision making. Economic uncertainty affect decision making in all layers of the economy with lowest income to highest income and from households to global firm owners. Moreover, EPU fluctuations not only make investors to react, but also banking community, managers, and politicians. Duo to the diversity of the parameters and areas that are affected by the EPU, we summarize a limited number of them in Table 2. We should note that the relation as well as the direction and the time horizons are enlisted with respect to the indicated references. The relations are unidirectional from EPU to variables. Although, there exist bidirectional relations, for the sake of simplicity, we address one side of the relations only.

| # | Variable | Relation/Direction (Short-/Long-run) $EPU \longmapsto Var$ | Reference (s) |
|---|----------|-----------------------------------------------------------|---------------|
| 1 | Stock Market Volatility | Positive (S), Negative (S, L) | (Choi & Shim, 2019; Li et al., 2019; Liu & Zhang, 2015; Ma et al., 2020; Manela & Moreira, 2017; Shaikh, 2019; Škrinjarić & Orlović, 2020; Xia et al., 2020; Yu et al., 2018) |
| 2 | Firms' Investment | Negative (S, L) | ( Baker et al., 2016; Chen et al., 2019; Feng et al., 2019; Gulen & Ion, 2016; Kang et al., 2014) |
| 3 | Household Investment | Negative (S, L) | (Da et al., 2015; Lee et al., 2020; Lunn & Kornrich, 2018; van der Wielen & Barrios, 2020) |
| 4 | Inflation | Positive, Negative (L) | (Balcilar et al., 2017; Colombo, 2013; Gao et al., 2019; Glas & Hartmann, 2016) |
| 5 | Cash Flow | Negative | (Ashraf & Shen, 2019; Belo & Yu, 2013; Liu et al., 2017; Liu et al., 2020) |
| 6 | Crash Risk | Positive | (Antonakakis et al., 2015; Arouri et al., 2016; Fang et al., 2018; Jin et al., 2019; Luo & Zhang, 2020) |
| 7 | Interest Rate (Average Interest Rate) | Positive (S, L) | (Ashraf & Shen, 2019; Baker et al., 2016; Bholat et al., 2015; Francis et al., 2014; Gholipour, 2019; Poza & Monge, 2020) |
| 8 | Loan Price | Positive (S) | (Ashraf & Shen, 2019; Chi & Li, 2017) |
| 9 | Capital Expenditure | Negative | (Akron et al., 2020; Kahle & Stulz, 2013; Tran, 2019; Wu et al., 2020) |
| 10 | Liquidity (e.g., Bank Liquidity, Stock Liquidity, and Capital Liquidity) | Positive (S, L) | (Chung & Chuwonganant, 2014; Dash et al., 2019; Debata & Mahakud, 2018; Gholipour, 2019; Huang & Luk, 2020) |
| 11 | Bitcoin returns | Positive (S), Negative (S) | (Demir et al., 2018; Fang et al., 2019; Mokni, 2021; Paule-Vianez et al., 2020; Qin et al., 2021; Shaikh, 2020) |

| # | Decision | Relation/Direction (Short-/Long-run) | Reference (s) |
|---|----------|--------------------------------------|---------------|
| 1 | Tax Avoidance | Positive (S) | (Kang & Wang, 2020; Nguyen & Nguyen, 2020; Shen et al., 2021) |
| 2 | Tax Burden | Positive (S) | (Dang et al., 2019; Duong et al., 2020) |
| 3 | Cash Holding (and Savings) | Positive (S, L) | (Caballero, 1990; Demir & Ersan, 2017; Feng et al., 2019; Giavazzi & McMahon, 2012; Im et al., 2017; Li, 2019; Liu & |

| | | | Zhang, 2020; Phan et al., 2019; Su et al., 2020) |
|---|---|---|---|
| 4 | Asset Pricing | Negative (S) | (Brogaard & Detzel, 2015; Kannadhasan & Das, 2020; Liu et al., 2017; Pastor & Veronesi, 2012) |
| 5 | Diversification (Corporate and Investment Diversification) | Positive | (Das et al., 2019; Hoang et al., 2021; Nguyen & Nguyen, 2020; Tran et al., 2020) |
| 6 | Earning Management | Positive<br>Negative | (Cui et al., 2020; Haque et al., 2019; Roma et al., 2020; Yiqiang Jin et al., 2019; Yung & Root, 2019) |
| 7 | Mergers & Acquisitions (M&A) | Positive (S),<br>Negative (S, L) | (Bonaime et al., 2018; Duchin & Schmidt, 2013; Kannadhasan & Das, 2020; Li et al., 2021; Nguyen & Phan, 2017; Sha et al., 2020) |
| 8 | Voluntarily Information Disclosure | Positive | (Guay et al., 2016; Nagar et al., 2019) |
| 9 | Innovation in R&D | Positive | (Al-Thaqeb et al., 2020; Gholipour, 2019; Li et al., 2019) |
| 10 | Bank Lending | Positive, Negative | (Alessandri & Bottero, 2020; Biswas & Zhai, 2021; Braun, 2020; Danisman et al., 2021; Hu & Gong, 2019) |

# 5    Conclusion and Future Works

Rising uncertainty widely affects economic development by putting a delay on investment. Different studies confirm that as the uncertainty grows, the amounts of investment shrink in firms (Bloom, 2009; Braun, 2020; Chen et al., 2020). Furthermore, results of past research works confirmed that uncertainty shocks lead to sharp recessions (Bloom, 2009, 2017; Caggiano et al., 2014; Fontaine et al., 2017), and Ercolani and Natoli (2020) showed that the recessions could be predicted using the uncertainty level. Economic policy uncertainty which is the uncertainty about monetary policies perceived by the investors, is a critical indicator in economic studies to predict future investments, the unemployment rate, and recessions. In this paper, we intended to review the economic policy uncertainty and surveys the EPU measurement techniques with a focus on text mining methods. We divide the EPU measures into three major groups with respect to the type of their input data. The first group uses financial data and tries to track the stock market's volatility as an indirect indicator of the investors' uncertainty. The second one adopts textual data with an assumption that the level of uncertainty of investors can be determined with respect to the news they receive. The third group of methods utilizes diverse types of information, e.g., google trend time series and marketing behaviors, to estimate the realized level of uncertainty. We summarize the pros and cons of these methods in Table 1 to compare different methods.

Methods of the second group are dominantly studied to estimate the uncertainty realized by the public community and firm owners through the news they receive. These methods enable us to extract the knowledge and sentiments hidden in the textual material are called text mining methods. Moreover, the hidden knowledge reliably demonstrates the situation of the economies with little transparency without having to deal with collecting sensitive financial information.

There are four various types of text mining methods that have been used to measure the economic policy uncertainty, that are (1) keyword-based, (2) topic modeling, (3) word embedding-based and (4) document classification. A keyword-based method proposed by Baker et al. (2016) is widely used and has shown its reliability. Baker et al. (2016) used a pre-defined list of words representing the "Policy," "Economy," and "Uncertainty" concepts. This keyword list makes their method depend on the quality and coverage of the list. Moreover, it leads to compatible but different results for countries with languages other than English. As we enlisted, several reported methods have been tested to overcome the obstacles of the Baker's et al. (2016) method.

To the best of our knowledge, no study reviews the main areas which are benefitted from the economic policy uncertainty measurement. We designed our article to cover a wide range of domains, including (1) health, (2) personal issues, (3) tourism, (4) real estate, (5) climate change, (6) geopolitical risk, and (7) general micro-/macro-economic areas, with the hope that studying the impacts of uncertainty would attract further attention of the researchers from various research fields. We believe future studies will continue in two directions, that are measuring the EPU index and investigating its impacts. We present our suggestions for each direction as follows,

- Measuring the EPU index
  - o Providing a collection of labeled documents

    Keith et al. (2020) addressed the annotation bias in the Baker's et al. (2016) work that might come from the annotators' presumptions, errors in labeling, or content ambiguity. Thus, a precisely-labeled repository of news articles expressing economic uncertainty would benefit the EPU measurement. Furthermore, labeled documents' repositories facilitate effective topic modeling using the transformer-based word embedding representations, such as BERT and ELMo.

- Group-wise uncertainty analysis

  During our investigations, we found time windows included uncertainty-imposed terms expressing more than one change in the economic policy of the country. The impacts of a new policy may exaggerate the effects of another altered policy. Therefore, it is reasonable to group the documents addressing a particular policy together. Then, the uncertainty indices of each group can be measured apart from others. A regression analysis of the panel data of uncertainty measures proxied by an exogenous variable such as stock market volatility can reveal different policies' impacts.

- Developing the EPU using social media data

  The recently proposed Twitter-based economic uncertainty (TEU) measure provides an opportunity to study economic consequences in a micro-scale and predicting households' investment behaviors. Therefore, it would be beneficial to measure the TEU and utilize it to predict hedging and stock market investment.

- Impacts of the EPU
  - Impacts of the EPU on the economic responses to the COVID-19 pandemic

    The COVID-19 pandemic calls for urgent actions to reduce the adverse impacts of the outbreak. Governments worldwide have taken immediate actions to support low-income groups and firms and reinforce the public health systems. Whereas these actions are taken to diminish the pandemic's negative impacts, the uncertainty about them may adversely influence the economy. Thus, we recommend that the EPU comes from the outcomes of the supportive policies to be studied.

  - Policy presentation recommendation

    Verbalizing a new policy and how it would be expressed in the media may affect the realized uncertainty. Consequently, a recommender system that predicts consequences of a political announcement and recommends the best verbalization with minimum negative impacts would be preferable.

  - Examining the impacts of the EPU on various areas using textual data

    Although the impacts of high EPU have been studied extensively within the economy, its socio-economic consequences have not been investigated. In fact, the influences of the EPU on cryptocurrency surges, tourism sentiment, public health satisfaction, terror attacks, etc., can be explored by mining publicly expressed sentiments in social media. We strictly recommend the computer science community examine this area and open a new horizon to the economic world.

**Declarations**

## Funding

This research did not receive any specific grant from funding agencies in the public, commercial, or not-for-profit sectors.

## Declaration of Competing Interest

The authors report no declarations of interest.

## Disclosure of conflicts of interest

None.

## Data/Materials Availability

This review does not include data.

## Code Availability

This study does not include coding analysis.

## CRediT authorship contribution statement

**F. Kaveh-Yazdy**: Conceptualization, Searching, Writing - Original Draft, Writing - Review & Editing, Project administration.

**Sajjad Zarifzadeh**: Validation, Writing - Original Draft, Writing - Review & Editing, Resources.